\documentclass[12pt]{article}
\usepackage{amsmath}
\usepackage[pctex32]{graphics}

\begin{document}

\title{\textbf{Bose-Einstein Condensation in the Relativistic Ideal Bose Gas}%
}
\author{M. Grether,$^{a}$ M. de Llano,$^{b}$ and George A. Baker, Jr.$^{c}$ 
\\
$^{a}$Facultad de Ciencias, Universidad Nacional Aut\'{o}noma de M\'{e}xico 
\\
04510 M\'{e}xico, DF, Mexico\\
$^{b}$Texas Center for Superconductivity, University of Houston, Houston, TX%
\\
77204, USA and Instituto de Investigaciones en Materiales\\
Universidad Nacional Aut\'{o}noma de M\'{e}xico, 04510 M\'{e}xico, DF,
Mexico* \\
$^{c}$Theoretical Division, Los Alamos National Laboratory\\
Los Alamos, NM 87545, USA}
\maketitle

\begin{abstract}
The Bose-Einstein condensation (BEC) critical temperature in a relativistic
ideal Bose gas of identical bosons, with and without the antibosons expected
to be pair-produced abundantly at sufficiently hot temperatures, is exactly
calculated for all boson number-densities, all boson point rest masses, and
all temperatures. The Helmholtz free energy at the critical BEC temperature
is \textit{lower }with antibosons, thus implying that omitting antibosons
always leads to the computation of a metastable state.

PACS \# 03.75.Kk; 05.30.Jp; 47.75.+f
\end{abstract}

Since its theoretical prediction by Einstein in 1925 based on the work in
1924 by Bose on photons, and after languishing for many decades as a mere
academic exercise in textbooks, BEC has been observed in the laboratory in
laser-cooled, magnetically-trapped ultra-cold bosonic atomic clouds of $%
_{37}^{87}$Rb \cite{Ander},$\ _{3}^{7}$Li \cite{Bradley},$\ _{11}^{23}$Na 
\cite{Davis},$\ _{1}^{1}$H \cite{Fried}, $_{37}^{85}$Rb \cite{Cornish}, \ $%
_{2}^{4}$He \cite{Pereira}, $_{19}^{41}$K \cite{Mondugno}, $_{55}^{133}$Cs 
\cite{Grimm}(a), $_{70}^{174}$Yb \cite{Grimm}(b), and $_{24}^{52}$Cr \cite{Griesmaier}. Previously, BEC in a gas
of excitons had also been reported \cite{excitonicBEC}.\ More recently, BEC
has been seen as well in fermionic atomic gases of $_{19}^{40}$K \cite{Jin}
and $_{3}^{6}$Li \cite{Zwierlein} as a result of some of the fermions
presumably Cooper-pairing \cite{cooper}\ into bosons. The role of \textit{%
hole }Cooper pairs accounted for or not, along with the usual particle
Cooper pairs, has been explored \cite{Tolma2000}-\cite{PC07}\ with striking
implications for many-electron superconducting states within a BEC scenario.
Their effects in neutral-fermion many-particle systems are yet to be
investigated. It is the analogy of hole pairs with antibosons which has
piqued our interest in the problem to be dealt with in this paper. Although
it is hard to imagine a many-boson system whose constituent bosons do not
disintegrate into their components at the high temperatures where antiboson
production becomes substantial, it may shed light on the dynamics of
many-fermion systems where both particle- and hole-Cooper-pairing can occur
at \textit{all} temperatures. Moreover, recent specific-heat measurements 
\cite{Sherman03}\ in $TlCuCl_{3}$ suggest a magnon BEC with a relativistic
dispersion relation.

\qquad In early papers \cite{Landsberg65}-\cite{Beckmann82}\ on the
relativistic ideal boson gas (RIBG) explicit BEC\ critical transition
temperature $T_{c}$-formulae were derived for both the nonrelativistic and
ultra-relativistic limits and specific-heat anomalies at $T_{c}$ were
studied. In addition, Refs. \cite{Beckmann79}\cite{Beckmann82} considered
all space dimensions $d>0$ and delved into the relation between $d$ and
various critical exponents. Antiboson production, however, was \textit{not}
accounted for. The first papers to include \textit{both} bosons and
anti-bosons appear to be Refs. \cite{HaberWeldon81}\cite{HaberWeldon82}
where high-temperature expansions for the various thermodynamic functions
(pressure, particle-number density, entropy, specific heats, etc.) were
derived. Extensive numerical work in $d$ dimensions that does not rely such
high-temperature expansions was reported in Refs. \cite{SinghPandita83}\cite%
{SinghPathria84}.\ In an elegant treatment \cite{Goulart89}\ with inverse
Mellin transforms the specific heat anomaly of the RIBG at its BEC $T_{c}$
was found to be washed out when pair-production was included. The
relationship between the BEC of the RIBG and spontaneous-symmetry breaking
was explored in Refs. \cite{HaberWeldon82}\cite{Kapusta81}; see also the
rather complete Ref. \cite{KapustaGale}, esp. Sec. 2.4. The so-called BCS%
\cite{BCS}-to-Bose crossover scenario (see Ref. \cite{PC07} and refs.
therein), and even the pseudogap concept \cite{Timusk99}\ of
superconductors, first seemed to have appeared in quark physics in Ref. \cite%
{Babaev00}; for a review see Ref. \cite{Babaev01}. More recently, \textit{%
two }\textquotedblleft crossovers\textquotedblright\ have been identified 
\cite{Nishida05} in an interacting \textit{fermion} gas where pairing into
bosonic Cooper pairs \cite{cooper}\ can occur to form a relativistic
superfluid as an example of the BCS-Bose crossover followed by a
Bose-to-RIBG/BEC-crossover where both anti-bosons as well as bosons dominate
the thermodynamics.\ A fully relativistic detailed study \cite{He07}\ of
these crossovers at zero temperature has also appeared.

\qquad In this Letter we exhibit, as a function of boson number density,
exact BEC transition temperatures for the RIBG gas of identical bosons with
and without antibosons in 3D. The system with both kinds of bosons always
has the higher $T_{c}$, i.e., is the system with the first BEC singularity
that appears as it is cooled. This suggests that the Helmholtz free energy
might be lower and thus correspond thermodynamically to the \textit{stable}
system as opposed to a \textit{metastable} system for the lower-$T_{c\text{ }%
}$system. It is then calculated and indeed found to be lower at all
densities for the complete problem with both bosons and antibosons, when
compared to the problem without antibosons. This implies that the omission
of antibosons will \textit{not}\ lead to stable states.

\qquad The number of bosons $N$ of mass $m$\ that make up an ideal boson
gas\ in $d$ dimensions (\textit{without} antibosons) is $N=\sum_{\mathbf{k}%
}n_{\mathbf{k}}\equiv \sum\limits_{\mathbf{k}}[e^{\beta \{\mid E_{k}\mid
-\mu (T)\}}-1]^{-1}$\ where $\beta =1/k_{B}T$, $k_{B}$ is the Boltzmann
constant and $\mu (T)$ is the boson chemical potential. Here, the total
energy of each boson is%
\begin{eqnarray}
\mid E_{k}\mid &\equiv &\sqrt{c^{2}\hbar ^{2}k^{2}+m^{2}c^{4}}
\label{Eabsolute} \\
&=&mc^{2}+\hbar ^{2}k^{2}/2m+O(k^{4})\ \ \ \ \ \ \ \ \ \ \ \ \text{if }%
c\hbar k\ll mc^{2}\text{ \ \ \ \ \ \ \ \ \ \ }\mathbf{NR}  \label{NR} \\
&=&c\hbar k[1+{\frac{1}{2}}(mc/\hbar k)^{2}+O(k^{-4})]\ \ \ \ \ \text{if\ }%
c\hbar k\gg mc^{2}\text{ \ \ \ \ \ \ \ \ \ \ }\mathbf{UR}  \label{UR}
\end{eqnarray}%
where $k$ is the boson wavenumber, $m$ its rest mass and $c$ is the speed of
light. The two limits refer to the nonrelativistic $(NR)$ and
ultrarelativistic $(UR)$ extremes. For a cubic box of side length $L$ in the
continuous limit the sums over the $d$-dimensional wavevector $\mathbf{k}$
become integrals as $\sum\limits_{\mathbf{k}}\rightarrow \left( L/2\pi
\right) ^{d}\int d^{d}k.$ At the BEC critical transition temperature $T_{c}$%
, $\mu (T_{c})=mc^{2}$ and the boson number density can be expressed as 
\begin{equation}
n\equiv \frac{N}{L^{d}}=\frac{1}{\left( 2\pi \right) ^{d}}\int d^{d}k\;\frac{%
1}{\exp \left[ \beta _{c}\left( \mid E_{k}\mid -\text{ }mc^{2}\right) \right]
-1}  \label{BB1}
\end{equation}%
where $\beta _{c}\equiv 1/k_{B}T_{c}$. In the nonrelativistic (NR) extreme (%
\ref{NR}) inserted into (\ref{BB1}) leaves

$n=\left( 2\pi \right) ^{-d}\int d^{d}k\ [\exp \left[ \beta _{c}\left( \hbar
^{2}k^{2}/2m\right) \right] -1]^{-1}$. Putting $d^{d}k=[2\pi ^{d/2}/\Gamma
(d/2)]k^{d-1}dk$ when integrating over terms independent of angles gives an
expression for $n$ as function of $T_{c}$\ in terms of the Bose function 
\cite{Path}\ $g_{\sigma }(z)$\ of $z\equiv \exp (\mu /k_{B}T_{c})$ which for 
$z=1$\ diverges, namely $g_{\sigma }(1)\rightarrow \infty $ when $\sigma
\leq 1$, but becomes the Riemann Zeta function $\zeta (\sigma )<\infty $
when $\sigma >1$. Here $\Gamma (\sigma )$ is the gamma function. Solving for
the critical temperature then gives 
\begin{equation}
k_{B}T_{c}^{NR\text{-}B}=\frac{2\pi \hbar ^{2}}{m}\left[ n/\zeta (d/2)\right]
^{2/d}  \label{NRd}
\end{equation}%
where the superscript $NR$-$B$ stands for the nonrelativistic limit with
bosons ($B$) but no antibosons ($\overline{B}$). In 3D this reduces to the
familiar textbook result $k_{B}T_{c}^{NR\text{-}B}\simeq 3.31\hbar
^{2}n^{2/3}/m$\ since $\zeta (3/2)\simeq 2.612.$ In the ultrarelativistic
(UR) extreme the leading term of (\ref{UR}) inserted into (\ref{BB1}) leads
to $T_{c}=0$ for all $d\leq 1$ since then $g_{d}(1)$ diverges. However, for $%
d>1$%
\begin{equation}
k_{B}T_{c}^{UR\text{-}B}=\left[ \frac{\hbar ^{d}c^{d}2^{d-1}\pi ^{d/2}\Gamma
(d/2)}{\Gamma (d)\zeta (d)}\right] ^{1/d}n^{1/d}  \label{TcURBD}
\end{equation}%
which in 3D becomes $k_{B}T_{c}^{UR\text{-}B}=\hbar c\pi ^{2/3}[n/\zeta
(3)]^{1/3}\simeq 2.017\hbar cn^{1/3}$ as $\zeta (3)\simeq 1.20206.$ In 2D $%
T_{c}^{UR\text{-}B}\neq 0$ unlike the common instance with quadratic
dispersion where $T_{c}$ vanishes because $g_{1}(1)$ diverges; specifically $%
k_{B}T_{c}^{UR\text{-}B}=\hbar c[2\pi n/\zeta (2)]^{1/2}\simeq 1.954\hbar
cn^{1/2}$ since $\zeta (2)=\pi ^{2}/6.$

\qquad At sufficiently high temperatures such that $k_{B}T\gg mc^{2}$
boson-antiboson pair-production occurs abundantly; this has been stressed by
Huang \cite{Huang}. The total energy $E_{k}$ of each particle always
satisfies $E_{k}^{2}=c^{2}\hbar ^{2}k^{2}+m^{2}c^{4}$ so that $E_{k}=\pm
\mid E_{k}\mid $ where $\mid E_{k}\mid $ is given by (\ref{Eabsolute}) and
with the $+$ sign referring to bosons and the $-$ sign to antibosons.
Instead of $N=\sum_{\mathbf{k}}n_{\mathbf{k}}$ the \textit{complete} number
equation is now \cite{HaberWeldon81} 
\begin{eqnarray}
N\text{ \ }-\text{ }\overline{N} &\equiv &\sum_{\mathbf{k}}\left( n_{\mathbf{%
k}}-\overline{n_{\mathbf{k}}}\right)   \label{Q-1} \\
&=&\sum\limits_{\mathbf{k}}\left[ \frac{1}{\exp \left[ \beta (\mid E_{k}\mid
-\text{ }\mu )\right] -1}-\frac{1}{\exp \left[ \beta (\mid E_{k}\mid +\text{ 
}\mu )\right] -1}\right]   \notag
\end{eqnarray}%
where $n_{\mathbf{k}},$ ($\overline{n_{\mathbf{k}}}$) is the average number
of{\LARGE \ }bosons (antibosons) in{\LARGE \ }the state of energy $\pm \mid
E_{k}\mid $, respectively, at a given temperature $T$ and $N$ ($\overline{N}$%
) is their respective total number at that temperature. Since $n_{\mathbf{k}%
},\;\overline{n_{\mathbf{k}}}>0$ for all $\mathbf{k}$ and $E_{0}=mc^{2}$,
the chemical potential must be bounded by $-mc^{2}\leq \mu \leq mc^{2}.$\
Instead of $N$ constant one must now impose\ the constancy of $N-\overline{N}
$ to extract the correct BEC critical\ temperature, say, $T_{c}^{B\overline{B%
}}$ referring to both bosons ($B$) and antibosons ($\overline{B}$).\ Since $%
\left\vert \mu (T_{c}^{B\overline{B}})\right\vert =mc^{2}$ (\ref{Q-1})
becomes%
\begin{equation}
n\equiv (N-\overline{N})/L^{d}=\frac{2\pi ^{d/2}}{\Gamma (d/2)(2\pi )^{d}}%
\int\limits_{0}^{\infty }dkk^{d-1}\frac{\sinh (\beta _{c}mc^{2})}{\cosh
\left( \beta _{c}\sqrt{c^{2}\hbar ^{2}k^{2}+m^{2}c^{4}}\right) -\cosh (\beta
_{c}mc^{2})}.  \label{exactBA}
\end{equation}%
This is an exact expression for the BEC $T_{c}$ of an\ ideal Bose gas at 
\textit{any }temperature as it includes both bosons and antibosons; it is
consistent with Eq. (13) of Ref. \cite{SinghPandita83}.

\qquad At low enough temperatures such that $k_{B}T_{c}\ll mc^{2}$\
antibosons can be neglected and (\ref{exactBA}) simplifies to, say, $%
T_{c}^{NR\text{-}B\overline{B}}$ which is precisely (\ref{NRd}) as expected.
In the opposite extreme $k_{B}T_{c}\gg mc^{2}$ (\ref{exactBA}) leads to the
limiting expression, say, 
\begin{equation}
k_{B}T_{c}^{UR\text{-}B\overline{B}}=\left[ \frac{\hbar ^{d}c^{d-2}\Gamma
(d/2)(2\pi )^{d}}{4m\pi ^{d/2}\Gamma (d)\zeta (d-1)}\right]
^{1/(d-1)}n^{1/(d-1)}  \label{URBBbar}
\end{equation}%
that sharply differs from (\ref{TcURBD}). In 3D this becomes $k_{B}T_{c}^{UR%
\text{-}B\overline{B}}=(3\hbar ^{3}c/m)^{1/2}n^{1/2}$, a result apparently
first reported in Ref. \cite{HaberWeldon81}.\ This novel relation has
suggested \cite{Sherman03}\ itself experimentally as a magnon BEC in
specific-heat measurements in $TlCuCl_{3}.$

\qquad As functions of the dimensionless boson number-density $\hbar
^{3}n/m^{3}c^{3}$ Fig. 1 displays the behavior of the exact $T_{c}^{B%
\overline{B}}$ (in units of $mc^{2}/k_{B}$) numerically extracted from (\ref%
{exactBA}) for $d=3$ (thick full curve labeled \textquotedblleft exact B$%
\overline{\text{B}}$\textquotedblright )\ compared with the nonrelativistic
limit $T_{c}^{NR\text{-}B\overline{B}}$ from (\ref{NRd}) (dashed line
labeled \textquotedblleft NR-B$\overline{\text{B}}$\textquotedblright )\ and
with the ultrarelativistic $T_{c}^{UR\text{-}B\overline{B}}$ just stated
(full thin line labeled \textquotedblleft UR-B$\overline{\text{B}}$%
\textquotedblright ).\ Fig. 2 shows how, at sufficiently high densities $n$
and/or sufficiently small boson rest mass $m,$\ the exact $T_{c}^{B\overline{%
B}}$ (again in units of $mc^{2}/k_{B}$, full curve labeled \textquotedblleft
exact B$\overline{\text{B}}$\textquotedblright ) is clearly the \textit{first%
} BEC singularity encountered as the many-boson system is cooled, \ compared
with the \textquotedblleft later\textquotedblright\ BEC singularity in the
system \textit{without} antibosons at $T_{c}^{B}$ (dashed curve labeled
\textquotedblleft exact B\textquotedblright ) extracted numerically from (%
\ref{BB1}). It is then tempting to speculate that the boson gas with \textit{%
both} kinds of bosons will be the more stable, i.e., have a \textit{lower }%
Helmholtz free energy at all critical temperatures, at any fixed $\hbar
^{3}n/m^{3}c^{3}$.\ This will now be shown to be the case indeed.

\qquad The exact Helmholtz free energy per unit volume $V=L^{3}$ for the
boson-antiboson 3D mixture, when $T=T_{c}\equiv 1/k_{B}\beta _{c}$ and $\mu
=mc^{2}$, is 
\begin{eqnarray}
F^{\text{exact}B\overline{B}}(T_{c},V)/V &=&nmc^{2}+(k_{B}T_{c}/2\pi
^{2})\int_{0}^{\infty }dkk^{2}\{\hbox{ln}\lbrack 1-\exp (\beta
_{c}[mc^{2}-\mid E_{k}\mid ])]  \notag \\
&&+\hbox{ln}\lbrack 1-\exp (-\beta _{c}[mc^{2}+\mid E_{k}\mid ])]\}
\label{exactF}
\end{eqnarray}
where $\mid E_{k}\mid $ is given by (\ref{Eabsolute}) and $n=(N-\overline{N}%
)/V$, and $T_{c}$ is extracted numerically from (\ref{exactBA}) for each
value of $n$. If $k_{B}T_{c}\,<<\,mc^{2}$ antibosons can be neglected
entirely and $F^{\text{exact}B}(T_{c},V)$ is just (\ref{exactBA}) but
without the second log term and for which $T_{c}$ is now extracted
numerically from (\ref{BB1}) instead of from (\ref{exactBA})\ for each value
of $n$. Figure 3 is the difference between the two free energies and more
clearly shows why $F^{\text{exact}B\overline{B}}(T_{c},V)$ is always lower.
The inset figure shows the behavior of these two free energies, each
evaluated at their appropriate $T_{c}$ value; by inspection, both curves
correspond as they should to \textit{positive} pressures $P$ since $%
P=-(\partial F/\partial V)_{N,T}$. Figure 3 proves the speculation advanced.

\qquad In summary, based on exact numerical calculations of BEC $T_{c}$s\ in
a 3D RIBG with and without the antibosons expected to be pair-produced at
higher and higher temperatures, the highest critical $T_{c}$ is that
associated with the RIBG system with \textit{both }kinds of bosons taken
into account. At lower $n$ and/or larger $m$ the higher $T_{c}$ merges
smoothly from above onto the lower $T_{c}$ of the RIBG system with
antibosons neglected. Comparing the associated Helmholtz free energies shows
that the RIBG with both kinds of bosons has \textit{lower }values for all $n$
and $m$ and thus substantiates the initial suspicion that the RIBG system
with no antibosons is metastable with respect to the one with both kinds of
bosons.

\begin{figure}[tbph]
\vspace{4.5cm}
\hspace{4.5cm}
\special{eps: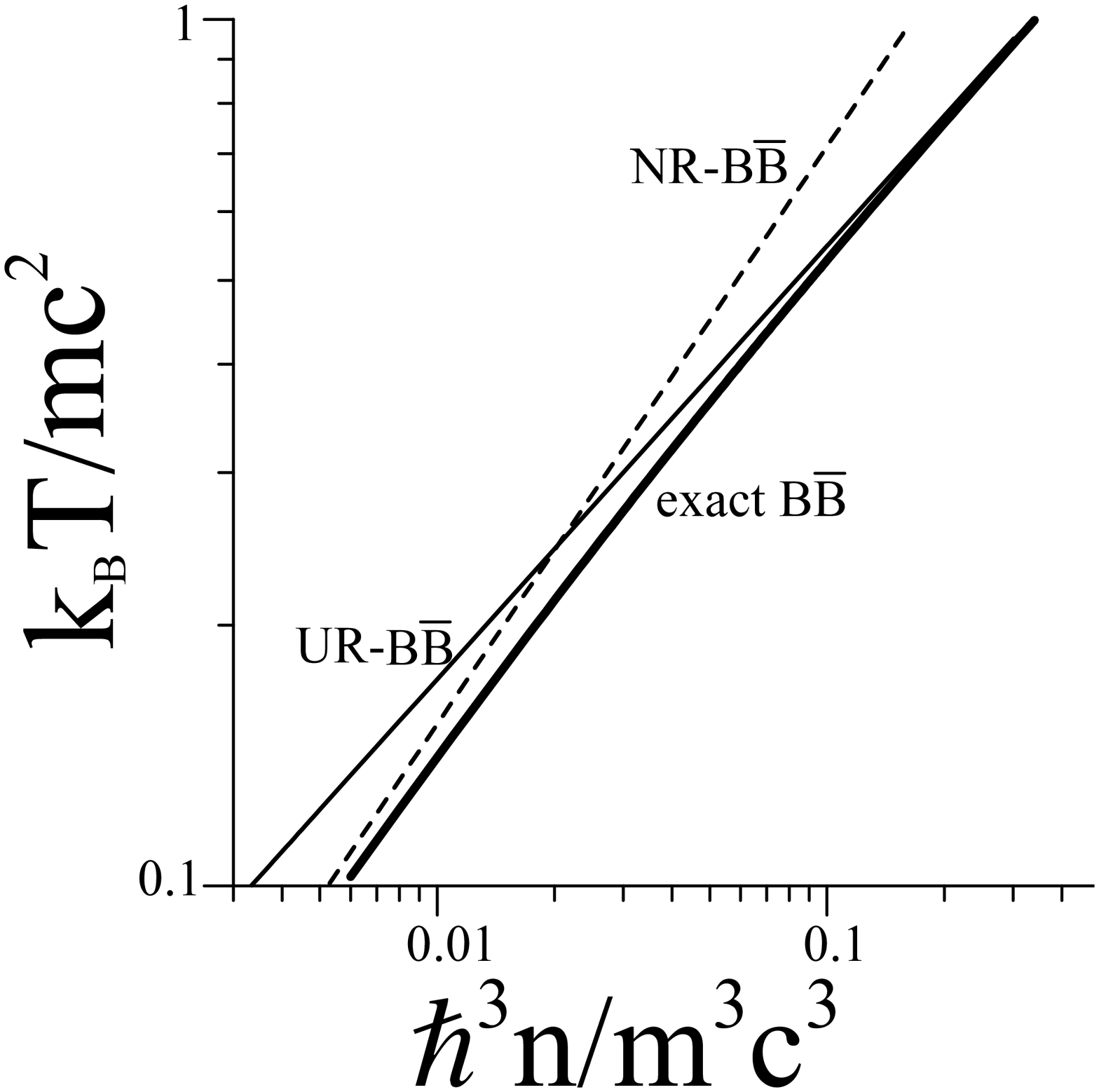  x=2.5in y=6cm}
\vspace{-0cm}
\begin{quote}
\noindent {\textbf{Figure 1.}} 
BEC $T_{c}$s (in units $mc^{2}/k_{B}$) as function of boson
number-density $n$ expressed in dimensionless form as $\hbar
^{3}n/m^{3}c^{3}.$ Thick curve labeled \textquotedblleft exact $B\overline{B}
$\textquotedblright\ is exact numerical result of (\ref{exactBA}) that
corresponds to BEC $T_{c}$\ in a RIBG with both bosons $B$\ \textit{and}
anti-bosons $\overline{B}$. Thin full straight line labeled
\textquotedblleft UR-$B\overline{B}$\textquotedblright\ is the
ultrarelativistic limit (\ref{URBBbar}) for $d=3$ with both kinds of bosons.
Dashed straight line labeled \textquotedblleft NR-$B\overline{B}$%
\textquotedblright\ is its corresponding nonrelativistic limit (\ref{NRd})
for $d=3$ and tends asymptotically to the \textquotedblleft exact $B%
\overline{B}$\textquotedblright\ curve at smaller $n$.
\end{quote}
\end{figure}

\begin{figure}[tbph]
\vspace{2.5cm}
\hspace{5cm}
\special{eps: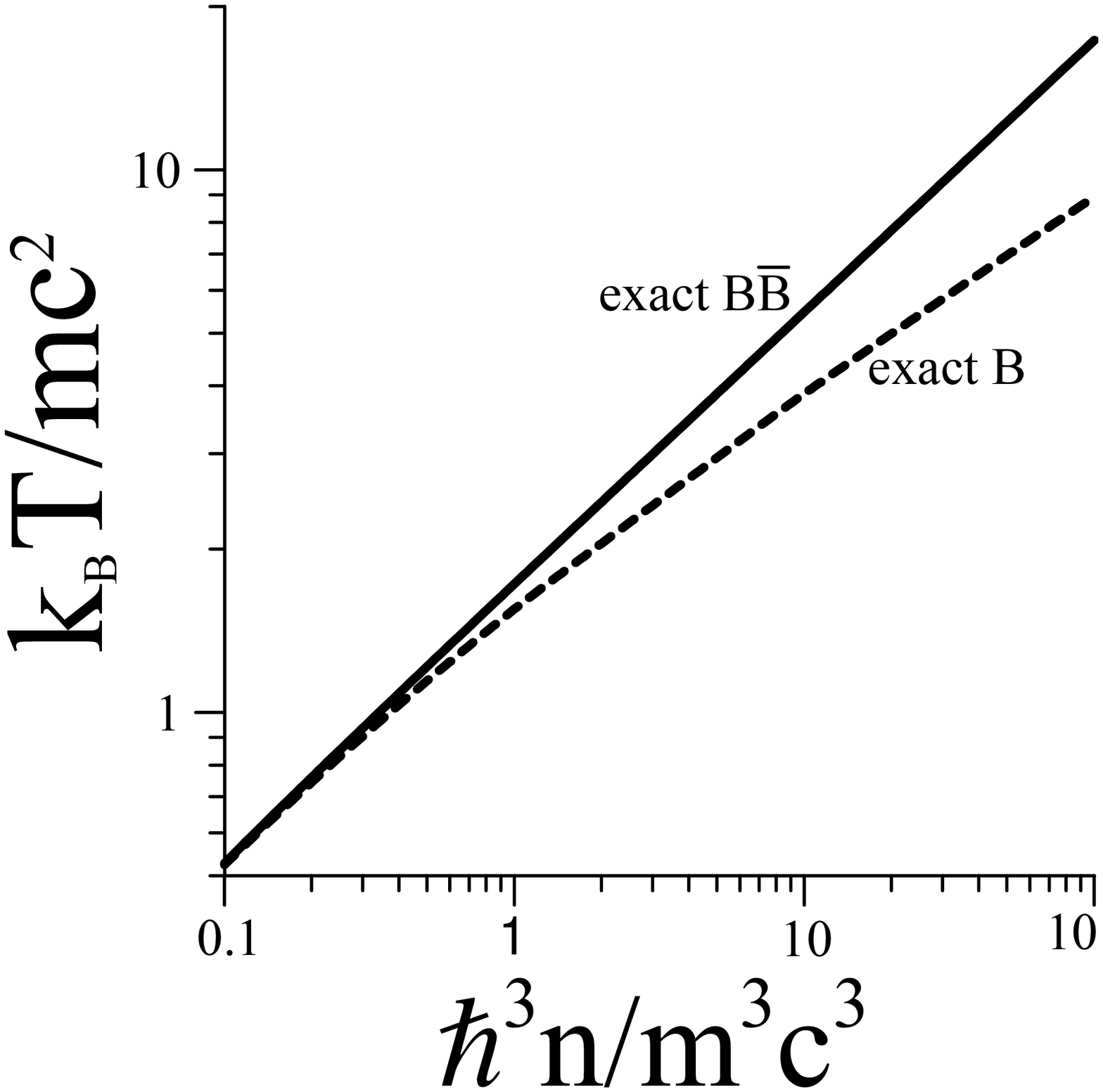  x=2.4in y=5.8cm}
\vspace{0cm}
\begin{quote}
\noindent {\textbf{Figure 2.}} Same as Fig. 1 comparing exact $B\overline{B}$ RIBG $T_{c}$\
extracted from (\ref{exactBA}) against that of exact $B$ from (\ref{BB1}).
\end{quote}
\end{figure}

\begin{figure}[tbph]
\vspace{6cm}
\hspace{4.5cm}
\special{eps: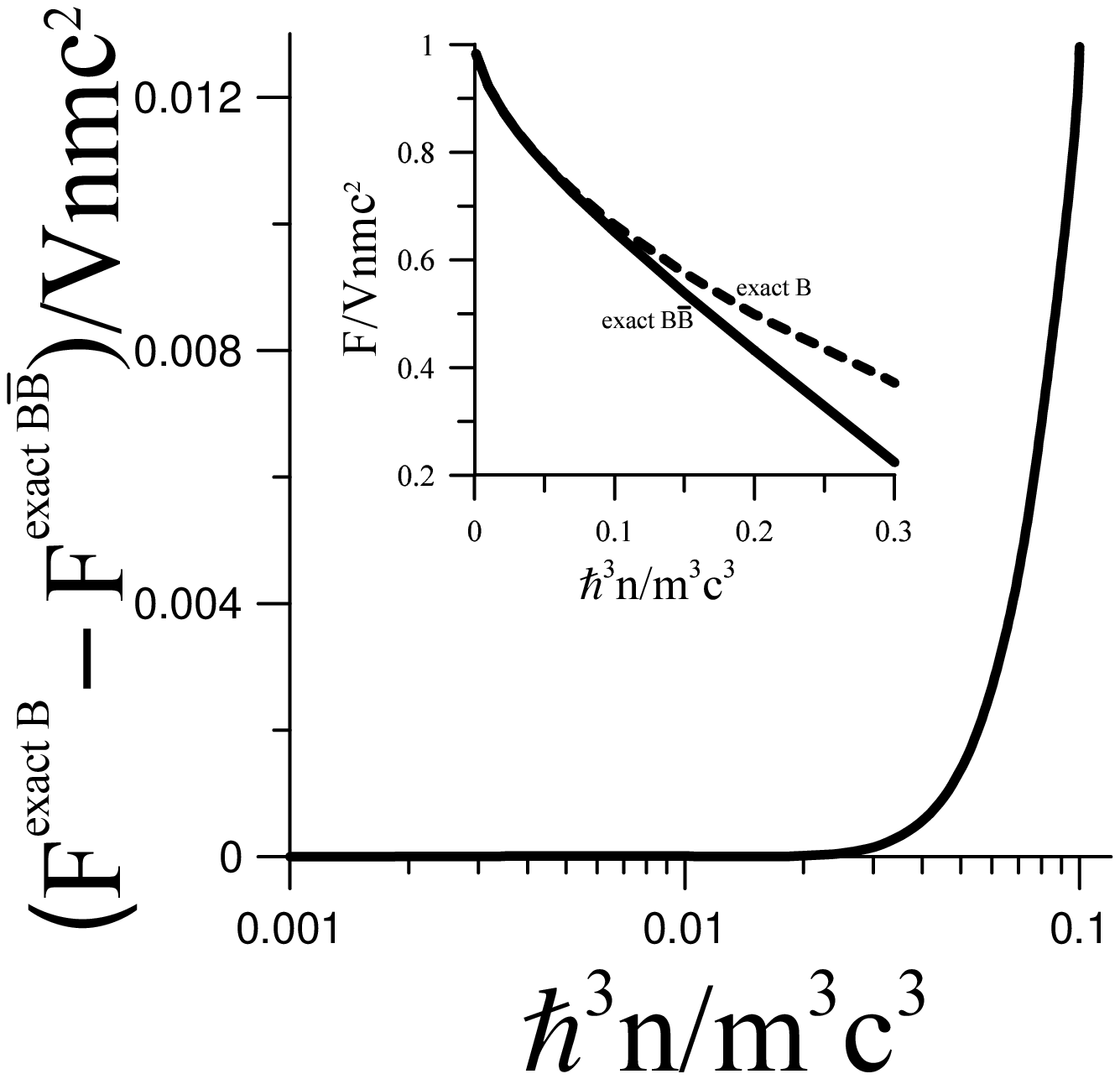  x=2.8in y=6.5cm}
\vspace{-0cm}
\begin{quote}
\noindent {\textbf{Figure 3.}} Difference between exact Helmholtz free energy density $F/V$\ (in
units of $nmc^{2}$ which is the total rest-mass energy density) (full curve
labeled \textquotedblleft exact$B\overline{B}$\textquotedblright ) and that
without anti-bosons (dashed curve labeled \textquotedblleft exact$B$%
\textquotedblright ) using same horizontal axes as in Figs. 1 and 2. The
inset figure displays the two free energies.
\end{quote}
\end{figure}
\textbf{Acknowledgments }MdeLl thanks E.H. Lieb for a discussion on the
contents of this paper and for his encouragement; he also acknowledges
support from UNAM-DGAPA-PAPIIT (Mexico) \#IN108205 and CONACyT (Mexico)
\#41302-F. The work of F.J. Sevilla in the initial phases of this research
is appreciated. We thank E.Ya. Sherman, L. He, and E. Babaev for calling
Refs. \cite{Sherman03}, \cite{Babaev00}\cite{Babaev01}, and \cite{He07},
respectively, to our attention.\ This work was supported in part by the U.
S. Energy Department at the Los Alamos National Laboratory under Contract \#
DE-AC 52-06NA25396.

*Permanent address.

\end{document}